\documentclass{article}
\usepackage{spconf,amsmath,graphicx,epsfig,epstopdf}

\title{Phonetic-attention scoring for deep speaker features \\ in speaker verification}
\name{Lantian Li, Zhiyuan Tang, Ying Shi, Dong Wang\thanks{
This work was supported by the National Natural Science
Foundation of China under Grant No. 61371136 and 61633013.
Dong Wang is the corresponding author (wangdong99@mails.tsinghua.edu.cn).}}
\address{CSLT, Tsinghua University, Beijing, 100084, China}%
\begin{document}
%
\maketitle
\begin{abstract}
Recent studies have shown that frame-level deep speaker features can be derived from a deep neural network with the training target
set to discriminate speakers by a short speech segment. By pooling the frame-level features, utterance-level representations,
called d-vectors,
can be derived and used in the automatic speaker verification (ASV) task.
This simple average pooling, however, is inherently sensitive to the phonetic content of the utterance.
An interesting idea borrowed from machine translation is the attention-based mechanism,
where the contribution of an input word to the translation at a particular time is weighted by an attention score.
This score reflects the relevance of the input word and the present translation. We can use the same idea to
align utterances with different phonetic contents.

This paper proposes a phonetic-attention scoring approach for d-vector systems. By this approach,
an attention score is computed for each frame pair. This score reflects the similarity of the two
frames in phonetic content, and is used to weigh the contribution of this frame pair in the utterance-based scoring.
This new scoring approach emphasizes the frame pairs with similar phonetic contents, which essentially provides a soft
alignment for utterances with any phonetic contents.
Experimental results show that compared with the naive average pooling, this phonetic-attention scoring approach can
deliver consistent performance improvement in ASV tasks of both text-dependent and text-independent.

\end{abstract}
\section{Introduction}
\label{sec:intro}

Automatic speaker verification (ASV) is an important biometric authentication technology and
has a broad range of applications. The current ASV approach can be categorized
into two groups: the statistical model approach and the neural model approach.
The most famous statistical models for ASV involve
the Gaussian mixture model-universal background model (GMM-UBM)~\cite{Reynolds00},
the joint factor analysis model~\cite{Kenny07} and the i-vector model~\cite{dehak2011front,Ioffe06,lei2014novel}.
As for the neural model approach, Ehsan et al. proposed the first successful implementation~\cite{ehsan14},
where frame-level speaker features were extracted from a deep neural network (DNN),
and utterance-level speaker representations (`d-vectors') were derived by
averaging the frame-level features, i.e., average pooling.
This work was followed by a bunch of researchers~\cite{liu2015deep,snyder2016deep,snyder2017deep,li2017deep}.

The neural-based approach is essentially a feature learning approach, i.e., learning
frame-level speaker features from raw speech. In previous
work, we found that by this feature learning, speakers can be discriminated by a speech segment as short
as $0.3$ seconds~\cite{li2017deep}, either a word or a cough~\cite{zhang2017deep}.
However, with the conventional d-vector pipeline, this brilliant frame-level
discriminatory power cannot be fully utilized by the utterance-level ASV, due to the simple average pooling.
This shortage was quickly identified by researchers, and hence almost all
the studies after Ehsan et al.~\cite{ehsan14} chose to learn representations of segments rather than frames,
the so-called \emph{end-to-end} approach~\cite{snyder2016deep,heigold2016end,zhang2016end,zhang2017end}. However, frame-level feature learning possesses its own
advantages in both generalizability and ease of training~\cite{wang2017deep}, and meets our long-term desire of
deciphering speech signals~\cite{li2018deep}. An ideal approach, therefore, is to keep the feature learning
framework but solve the problem caused by average pooling.

To understand the problem of average pooling, first notice that feature pooling is equivalent to score pooling.
To make the presentation clear, we consider the simple inner product score:

\vspace{-2.4mm}
\[
\vec{s}_u \cdot \vec{s}_{u'}= \frac{1}{|u|}\sum_{f \in u} \vec{v}_f \cdot \frac{1}{|u'|}\sum_{f' \in u'} \vec{v}_{f'},
\]
\vspace{-2.4mm}

\noindent where $u$ and $u'$ are two utterances in test, $f$ denotes frames; $\vec{v}_f$ and $\vec{s}_u$ are frame-level speaker features
and utterance-level d-vectors, respectively. A simple arrangement leads to:

\vspace{-2.4mm}
\[
\vec{s}_u \cdot \vec{s}_{u'}= \frac{1}{|u|}\frac{1}{|u'|}\sum_{f \in u} \sum_{f' \in u'} \vec{v}_f \cdot  \vec{v}_{f'}.
\]
\vspace{-2.4mm}

\noindent This formula indicates that with average pooling, the utterance-level score $\vec{s}_u \cdot \vec{s}_{u'}$ is the average
of the frame-level scores $\vec{v}_f \cdot  \vec{v}_{f'}$. Most importantly, the scores of all the frame pairs $(f,f')$ are equally weighted, which is
obviously suboptimal, as the reliability of scores from different frame pairs may be substantially different. In particular, a
pair of frames in the same phonetic context may result in a much more reliable frame-level score compared to a pair in different
phonetic context, as demonstrated by the fact that text-dependent ASV generally outperforms text-independent ASV. This indicates that
a key problem of the average pooling method is that phonetic variation may cause serious performance degradation. This partly explains
why d-vector systems are mostly successful in text-dependent tasks.

A simple idea is to discriminate frame pairs in similar / different phonetic contents,
and put more emphasis on the frame pairs in similar phones.
This can be formulated by:

\vspace{-2.4mm}
\begin{equation}
\label{eq:base}
\vec{s}_u \cdot \vec{s}_{u'}= \frac{1}{|u|}\frac{1}{|u'|}\sum_{f \in u} \sum_{f' \in u'} \alpha(f,f') \cdot \vec{v}_f \cdot  \vec{v}_{f'},
\end{equation}
\vspace{-2.4mm}

\noindent where $\alpha(f,f')$  represents the weight for the frame pair $(f,f')$, computed from the similarity of their phonetic contents.
This is essentially a soft-alignment approach that aligns two utterances with respect to phonetic contents, where $\alpha(f,f')$ represents the
alignment degree of frames $f$ and $f'$, derived from the phonetic information of the two frames.

The idea of soft-alignment was motivated by the \emph{attention mechanism}
in neural machine translation (NMT)~\cite{Bahdanau:2014},
where the contribution of an input word to the translation at a particular time is weighted by an attention score,
and this attention score reflects the relevance of the input word and the present translation.
We therefore name our new scoring model by Eq.~(\ref{eq:base}) as \emph{phonetic-attention scoring}.
By paying more attention to frame pairs in similar phonetic contents, this new scoring approach essentially
turns a text-independent task to a text-dependent task, hence partly solving the problem
caused by phone variation with the naive average pooling .

In the next section, we will briefly describe the attention mechanism. The phonetic-attention scoring approach will be presented in Section~\ref{sec:phonetic},
and the experiments will be reported in Section~\ref{sec:exp}. The entire paper will be concluded in Section~\ref{sec:cond}.

\vspace{-1mm}
\section{Attention mechanism}
\label{sec:attention}
\vspace{-1mm}

\begin{figure}[htb]
    \centering
    \includegraphics[width=0.85\linewidth]{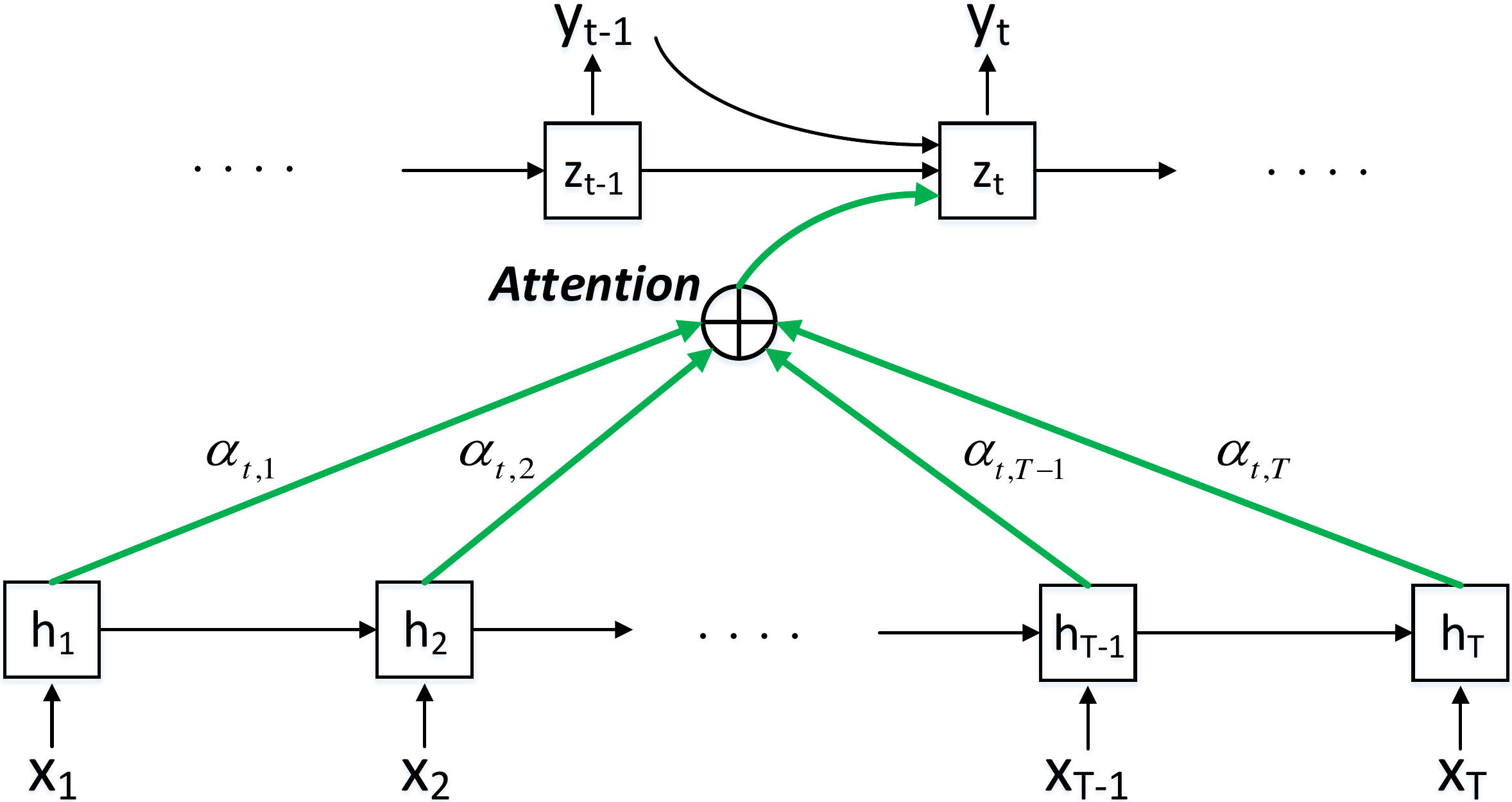}
    \caption{Attention mechanism in sequence to sequence model.}
    \vspace{-2mm}
    \label{fig:s2s}
\end{figure}

The attention mechanism was firstly proposed by~\cite{Bahdanau:2014} in the framework of sequence to sequence learning, and was applied to NMT.
Recently, this model has been widely used in many sequential learning tasks, e.g.,
speech recognition~\cite{chan2016listen}. In a nutshell, the attention approach looks up all the input elements (e.g., words in a sentence or frames in an utterance)
at each decoding time, and computes an attention weight for each element that reflects the relevance of that element with the present decoding. Based on these attention
weights, the information of the input elements is collected and used to guide decoding.
As shown in Fig.~\ref{fig:s2s}, at decoding time $t$, the attention weight $\alpha_{t,i}$ is computed for each input element $\vec{x}_i$ (more precisely, the annotation of $\vec{x}_i$, denoted by $\vec{h}_i$),
formally written as:

\[
\alpha_{t,i} = \sigma(g(\vec{z}_{t-1}, \vec{h}_i))
\]

\noindent where $\vec{z}_{t-1}$ is the decoding status at time $t$, and $g$ is a \emph{value function} that can be in any form. $\sigma$ is a normalization function (usually softmax) that ensures
$\sum_i \alpha_{t,i}=1$. The decoding for $\vec{y}_t$ is then formally written as:

\vspace{-2mm}
\[
\vec{y}_t = g'(\vec{z}_{t-1}, \vec{y}_{t-1}, \sum_i \alpha_{t,i}\vec{h}_i),
\]
\vspace{-2mm}

\noindent where $g'$ is the decoding model.
In the conventional setting, $g$ is a parametric function, e.g., a neural net, whose parameters are jointly optimized with other parts of the model, e.g., the decoding model $g'$.

\vspace{-1mm}
\vspace{-2mm}

\section{Phonetic-attention scoring}
\label{sec:phonetic}

\begin{figure}[htb]
    \centering
    \includegraphics[width=0.9\linewidth]{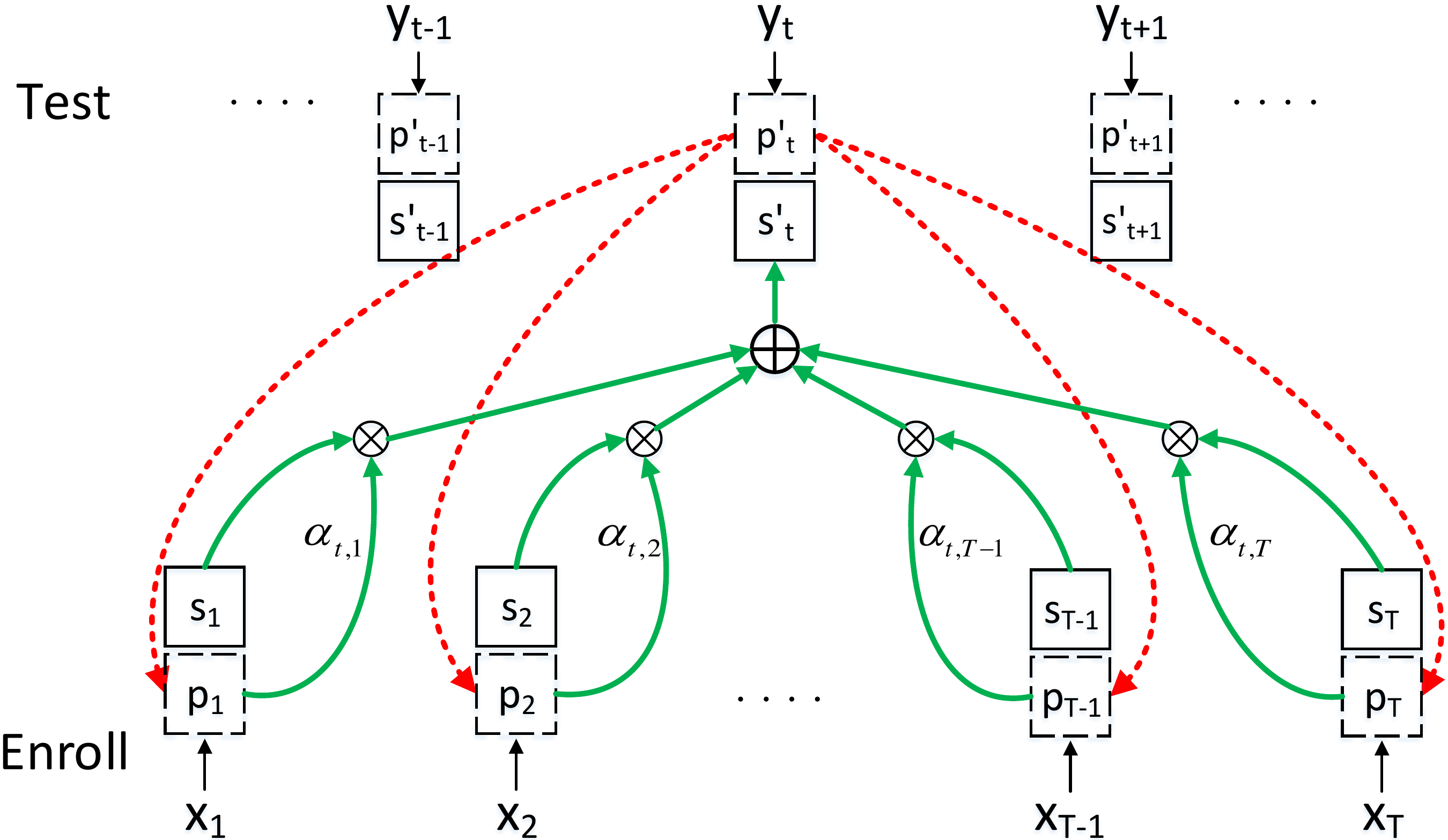}
    \caption{Diagram of the phonetic-attention model.}
    \label{fig:attention}
\end{figure}

We borrow the architecture shown in Fig.~\ref{fig:s2s} to build our phonetic-attention model in Eq.~(\ref{eq:base}).
Since our purpose is to align two existing sequences rather than sequence to sequence generation,
the structure can be largely simplified.
For example, the recurrent connection in both the input and output sequence can be omitted.
Secondly, in Fig.~\ref{fig:s2s}, the value function $g$ is learned from data; for our scoring model, we have a
clear goal to align utterances by phonetic content, so we can design the value function by hand (although
function learning with prior may help). This leads to the phonetic-attention model shown in Fig.~\ref{fig:attention}.

The architecture and the associated scoring method can be summarized into the following four steps:
\vspace{-2.4mm}

\noindent (1) For both the enrollment and test utterances, compute the frame-level speaker features from a speaker recognition DNN, denoted by
$S = [\vec{s}_1, \vec{s}_2, ... , \vec{s}_T]$ and $S' = [\vec{s}'_1, \vec{s}'_2, ... , \vec{s}'_{T'}]$.
Additionally, compute the frame-level phonetic features from a speech recognition DNN, denoted by
$P = [\vec{p}_1, \vec{p}_2, ... , \vec{p}_T]$ and $P' = [\vec{p}'_1, \vec{p}'_2, ... , \vec{p}'_{T'}]$.

\noindent(2) For each frame $t$ in the test utterance, compute the attention weight $\alpha_{t,i}$ for each frame $i$ in the enrollment utterance.

\vspace{-2mm}
\[
\alpha_{t,i} = \frac{\emph{KL}^{-1}(p'_t, p_i)}{\sum_i \emph{KL}^{-1}(p'_t, p_i)},
\]
\vspace{-2mm}

\noindent where the \emph{KL}$^{-1}(\cdot, \cdot)$ denotes the reciprocal of \emph{KL} distance. This step is represented by the red dashed line in Fig.~\ref{fig:attention}.

\noindent(3) Compute the matching score of frame $t$ in the test utterance as follows:

\vspace{-2mm}
\[
 d_t= \sum_i \alpha_{t,i} \cdot cos(s'_t, s_i).
\]
\vspace{-2mm}

\noindent This step is represented by the green solid line in Fig.~\ref{fig:attention}.

\noindent(4) Compute the matching score of the two utterances by averaging the frame-level matching score:

\vspace{-3.5mm}
\[
 d = \frac{1}{T} \sum_t d_t = \frac{1}{T} \sum_t \sum_i \alpha_{t,i} \cdot cos(s'_t, s_i).
\]
\vspace{-4mm}

\section{Related work}
\label{sec:attention}
\vspace{-1mm}

The attention mechanism has been studied by several authors in ASV, e.g.,~\cite{zhang2016end,rahman2018attention,liu2018exploring}.
However, most of the proposals used the attention mechanism to produce a better frame pooling, while we use it to produce a better
utterance alignment. In essence, these methods learn which frame should contribute to the speaker embedding, while our approach
learn which frame-pair should contribute to the matching score.
Moreover, most of these studies do not use phonetic knowledge explicitly, except~\cite{zhang2016end}.

Another work relevant to ours is the segmental dynamic time warping (SDTW) approach proposed by Mohamed et al.~\cite{adel2018text}.
This work holds the same idea as ours in aligning frame-level speaker features, however their alignment
is based on local temporal continuity, while ours is based on global phonetic contents.

\vspace{-1mm}
\section{Experiments}
\label{sec:exp}
\vspace{-1mm}
\subsection{Data}

\subsubsection{Training data}

The data used to train the d-vector systems is the \emph{CSLT-7500} database, which was collected by CSLT@Tsinghua University.
It consists of $7,500$ speakers and $1,532,766$ utterances. The sampling rate is $16$ kHz and the precision is $16$-bit.
Data augmentation is applied to cover more acoustic conditions, for which the MUSAN corpus~\cite{musan2015} is used to
provide additive noise, and the room impulse responses (RIRS) corpus~\cite{ko2017study} is used to generate reverberated samples.

\vspace{-2mm}
\subsubsection{Evaluation data}
\vspace{-1mm}

(1)~\emph{CIIH}: a dataset contains short commands used in the intelligent home scenario.
It contains recordings of $10$ short commands from $100$ speakers, and each command consists of $2$$\sim$$5$ Chinese characters.
For each speaker, every command is recorded $15$ times, amounting to $150$ utterances per speaker.
This dataset is used to evaluate the text-dependent (TD) task.

\noindent(2)~\emph{DSDB}: a dataset involving digital strings. It contains $1,099$ speakers, each speaking $15$$\sim$$20$ Chinese
digital strings. Each string contains $8$ Chinese digits, and is about $2$$\sim$$3$ seconds.
For each speaker, $5$ utterances are randomly sampled as enrollment, and the rest are used for test.
This dataset is used to evaluate the text-prompted (TP) task.

\noindent(3)~\emph{ALI-WILD}: a dataset collected by the Ali crowdsource platform.
It covers unlimited real-world scenarios, and contains $669$ speakers and $27,861$ speech segments.
We designed two test conditions: a short-duration scenario Ali(S) where the duration of the enrollment is $15$ seconds and the test is $3$ seconds,
and a long-duration scenario Ali(L) where the duration of the enrollment is $30$ seconds and the test is $15$ seconds.
This dataset is used to evaluate the text-independent (TI) task.

\vspace{-3mm}
\subsection{Settings}
\vspace{-1mm}

The DNN model to produce frame-level speaker features is a 9-layer time-delay neural network (TDNN), where the slicing parameters are
\{$t$-$2$, $t$-$1$, $t$, $t$+$1$, $t$+$2$\}, \{$t$-$2$, $t$+$2$\}, \{$t$\}, \{$t$-$1$, $t$+$1$\}, \{$t$\}, \{$t$-$2$, $t$+$2$\}, \{$t$\}, \{$t$-$4$, $t$+$4$\}, \{$t$\}.
Except the last hidden layer that involves $400$ neurons, the size of all other layers is $1,000$.
Once the DNN has been fully trained, $400$-dimensional deep speaker features were extracted from
the last hidden layer. The model was trained using the Kaldi toolkit~\cite{povey2011kaldi}. Based on this model,
we built a standard d-vector system with the naive average pooling, denoted by \emph{Baseline}.

The phonetic-attention model requires frame-level phonetic features.
We built a DNN-HMM hybrid system using Kaldi following the WSJ S5 recipe.
The training used $500$ hours of Chinese speech data.
The model is a TDNN, and each layer contains $512$ nodes.
The output layer contains $463$ units, corresponding to the number of GMM senones.
Once the model was trained, $463$-dimensional phone posteriors were derived from the output layer and were used as phonetic features.
The phonetic-attention system based on the phone posteriors is denoted by \emph{Att-Post}.
Another type of phonetic features can be derived from the final affine layer. To compress the size of the feature vector,
the Singular Value Decomposition (SVD) was applied to decompose the final affine matrix into
two low-rank matrices, where the rank was set to $100$. The $100$-dimensional activations were read from the low-rank layer
of the decomposed matrix, which we call bottleneck features. The phonetic-attention system based on the bottleneck features
is denoted by \emph{Att-BN}.

Finally, we built a phone-blind attention system where the attention weight is computed from the speaker feature itself, rather than
phonetic features. This approach is similar to the work in~\cite{rahman2018attention,liu2018exploring}, though the attention function is
not trained. This system is denoted by \emph{Att-Spk}.

\vspace{-4mm}
\subsection{Results}
\vspace{-2mm}

The results in terms of the equal error rate (EER) are shown in Table~\ref{tab:result}, where
the baseline system is based on the naive average pooling, while the three attention-based systems
use attention models based on different features. For each system, it reports results with two
frame-level metrics: cosine distance and cosine distance after LDA.
The LDA model was trained on CSLT-7500, and the dimensionality of its projection space was set to $150$.
There are four tasks in total:
the TD task on CIIH, the TP task on DSDB, the TI short-duration task on
Ali(S), and the TI long-duration task on Ali(L).
The best performance is marked in bold face.

From these results, it can be seen that on all these tasks, the attention-based systems outperform the
baseline system, indicating that the naive average pooling is indeed problematic. When comparing these
three attention-based systems, we find they perform quite different on different tasks. On the
TD task CIIH and TP task DSDB, the phone-blind attention system Att-Spk seems slightly
superior, while on the TI task Ali(S) and Ali(L), the two phonetic-attention systems are clearly better.
This observation is understandable, as on the TD or the TP tasks, the phonetic variation in
enrollment and test utterances are largely identical, so the appropriate alignment can be easily found by
even a phone-blind attention. On the TI tasks, however, the phonetic variation is much more
complex, for which additional phonetic information is required to align the enrollment and test utterances.
Finally, comparing the two phonetic-attention systems, the Att-BN is consistently better.
This indicates that the bottleneck feature is a more compact representation for the phonetic content.

\vspace{-5.5mm}
\begin{table}[htb!]
 \begin{center}
  \caption{Performance of different systems on different tasks.}
   \label{tab:result}
   \scalebox{0.9}{
     \begin{tabular}{|l|c|c|c|c|c|c|c|c|c|}
       \hline
             Systems           &   Metric     & \multicolumn{4}{|c|}{EER(\%)} \\
       \hline
                               &              & CIIH & DSDB & Ali(S) & Ali(L)\\
       \hline
              Baseline         &    Cosine    & 3.71 &  1.02  & 9.24& 4.95\\
                               &    LDA       & 2.49 &  0.70  & 5.84& 2.44\\
       \hline
             Att-Spk           &    Cosine    & 3.27 &  0.95  & 9.07& 4.95\\
                               &    LDA       & \textbf{2.11} &  \textbf{0.65}  & 5.80& 2.50\\
       \hline
             Att-Post          &    Cosine    & 3.28 &  0.97  & 9.12& 4.85\\
                               &    LDA       & 2.22 &  0.69  & 5.76& 2.32\\
       \hline
             Att-BN            &    Cosine    & 3.20 &  0.98  & 9.11&4.84\\
                               &     LDA      & 2.18 &  0.70  & \bf{5.69}&\bf{2.31}\\
       \hline
     \end{tabular}
   }
 \end{center}
\end{table}
\vspace{-12mm}

\subsection{Analysis}
\vspace{-2mm}

\begin{figure}[htb]
    \centering
    \includegraphics[width=0.9\linewidth]{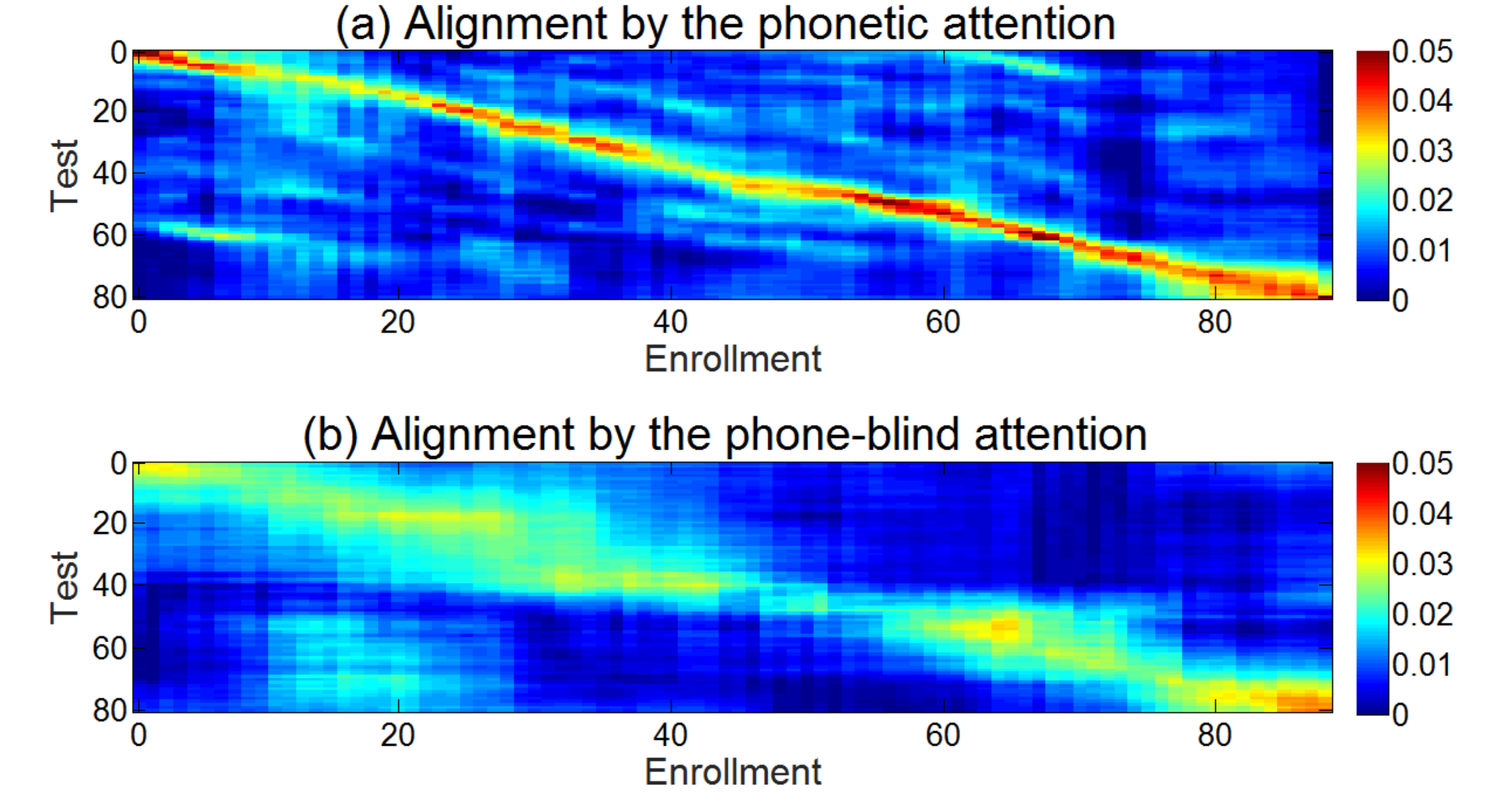}
    \vspace{-3.5mm}
    \caption{Alignment produced by the phone-blind and phonetic attentions on the TD task.}
    \label{fig:td}
\end{figure}

\begin{figure}[htb]
    \centering
    \includegraphics[width=0.9\linewidth]{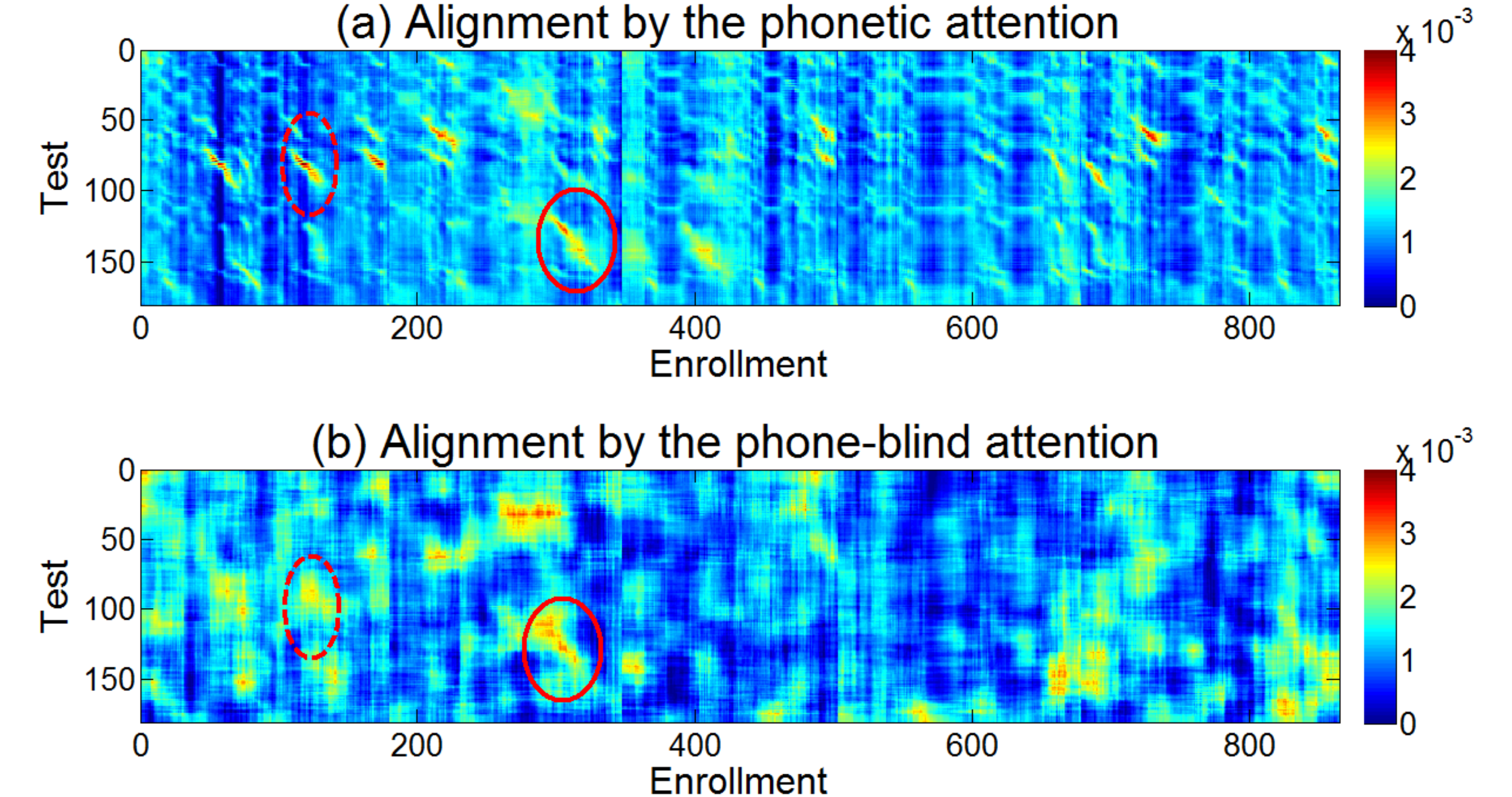}
    \vspace{-3.5mm}
    \caption{Alignment produced by the phone-blind and phonetic attentions on the TI task.}
    \label{fig:ti}
\end{figure}

To better understand the difference behavior of the phone-blind attention and the phonetic attention, we draw the alignment produced by them on
two samples from the TD and TI tasks respectively.
The figures are shown in Fig.~\ref{fig:td} and Fig.~\ref{fig:ti}.\footnote{The observations of the TD and TP tasks are quite similar, so here the figure on the TP task is omitted.}
It can be seen that on the TD task, two attention approaches produce similar alignments, while the alignment produced by
phonetic attention is more concentrated. This is not surprising, as the phonetic features are short-term and change more quickly than the speaker features.
Actually, this might be a key problem of the present implementation of the phonetic attention, as the concentration means less frames in one utterance being aligned for each frame in the other utterance, leading to unreliable scores.
Nevertheless, the explicit phonetic information does provide much more accurate alignments in the TI scenario, where the phonetic variation is
complex and phone-blind attention may produce rather poor alignments. This can be seen from Fig.~\ref{fig:ti} that the aligned segments produced by the
phonetic attention show clear slopped patterns, which is more realistic than the flat patterns produced by the phone-blind attention.

\vspace{-2mm}

\vspace{-3mm}

\vspace{-6mm}
\section{CONCLUSIONS}
\label{sec:cond}
\vspace{-2mm}

This paper proposed a phonetic-attention scoring approach for the d-vector speaker recognition system. This approach
uses frame-level phonetic information to produce a soft alignment between the enrollment and test utterances,
and computes the matching score by emphasizing the aligned frame pairs.
We tested the method on text-dependent, text-prompted and text-independent tasks, and found that it
delivered consistent performance improvement over the baseline system. The phonetic attention was also
compared with a naive phone-blind attention, and the results showed that the phone-blind attention worked well in
text-dependent and text-prompt tasks, but failed in text-independent tasks. Analysis was conducted to explain the observation.
In the further work, we will study speaker features that change more slowly. e.g., vowel-only feature.
It is also interesting to learn the value function.

\vfill\pagebreak

\bibliographystyle{IEEEbib}
\bibliography{refs}

\end{document}